\documentclass[prd,preprint,superscriptaddress,showpacs,byrevtex]{revtex4}
\usepackage{bm}
\def\fsl#1{\setbox0=\hbox{$#1$}           
   \dimen0=\wd0                                 
   \setbox1=\hbox{/} \dimen1=\wd1               
   \ifdim\dimen0>\dimen1                        
      \rlap{\hbox to \dimen0{\hfil/\hfil}}      
      #1                                        
   \else                                        
      \rlap{\hbox to \dimen1{\hfil$#1$\hfil}}   
      /                                         
   \fi}                                         %
\newcommand{\be}{\begin{equation}}
\newcommand{\ee}{\end{equation}}
\newcommand{\bea}{\begin{eqnarray}}
\newcommand{\eea}{\end{eqnarray}}
\newcommand{\beq}{\begin{equation}}
\newcommand{\eeq}{\end{equation}}
\newcommand{\beqs}{\begin{eqnarray}}
\newcommand{\eeqs}{\end{eqnarray}}

\begin{document}
\title{ Lattice QCD Method To Study Hadron Mass Is Not Correct }
\author{Gouranga C Nayak }\thanks{E-Mail: nayakg138@gmail.com}\thanks{G. C. Nayak was affiliated with C. N. Yang Institute for Theoretical Physics in 2004-2007.}
\affiliation{ C. N. Yang Institute for Theoretical Physics, Stony Brook University, Stony Brook NY, 11794-3840 USA}
\date{\today}
\begin{abstract}
Since the numerical path integration in the lattice QCD involves quark and gluon fields (not hadron fields) the lattice QCD cannot calculate any hadronic observable. Because of this reason the hadronic properties are extracted in the lattice QCD method by inserting complete set of hadron states $\sum_n |n><n|=1$ in between the partonic operators by assuming $H_{\rm QCD}|n>=E_n|n>$ where $E_n$ is the energy of the hadron. However, in this paper we find $H_{\rm QCD}|n>\neq E_n|n>$ because the QCD hamiltonian $H_{QCD}$ is unphysical but the $E_n$ and $|n>$ of the hadron are physical. We show that this is consistent with $E_{\rm QCD}(t)=<n|H_{QCD}|n>\neq E_n$ due to non-zero energy flux $E_{\rm flux}(t)$ in QCD because of confinement involving non-perturbative QCD. Hence we find that the lattice QCD method to study hadron mass is not correct.
\end{abstract}
\pacs{12.38.Aw, 12.38.Gc, 12.38.Lg, 11.30.Cp }
\maketitle
\pagestyle{plain}

\pagenumbering{arabic}

\section{Introduction}

The hadron (such as proton and neutron) is not an elementary particle of the nature but the quark and gluon inside the hadron are the elementary particles of the nature.
The interaction between quarks and gluons is caused by the color force or the strong force which is a fundamental force of the nature. Similar to Maxwell theory which describes the electromagnetic force of the nature, the Yang-Mills theory \cite{lym} describes the color force (or the strong force) of the nature. The quantum field theory of the classical Yang-Mills theory is the quantum chromodynamics (QCD) which describes the interaction between quarks and gluons.

Renormalization of non-abelian gauge theory was proved by 't Hooft and Veltman \cite{ltv} which enabled us to do the practical calculation in QCD. The discovery of the asymptotic freedom by Gross, Wilczek and Politzer \cite{lgw,lpo} proved that the renormalized QCD coupling decreases at small distance. In addition to renormalization the factorization theorem is important to study physical observable at high energy colliders \cite{lns}.

Due to the asymptotic freedom in QCD the short distance partonic level scattering cross section in the renormalized QCD can be calculated by using the perturbative QCD (pQCD). Hence there has been lot of progress in the pQCD calculation of the partonic level scattering cross section at LO, NLO, NNLO etc. at the high energy colliders.

Irrespective these progress in pQCD, since we have not directly experimentally observed quarks and gluons, the partonic level scattering cross section calculated by the pQCD can not be directly experimentally measured. What is directly experimentally measured is the hadron cross section. Hence it is necessary to know how the quarks and gluons form the hadron.

Since the asymptotic freedom predicts that the renormalized QCD coupling increases at long distance the pQCD is not reliable at long distance. Hence the non-perturbative QCD is necessary to study how the hadron is formed from quarks and gluons.

However, the analytic solution of the non-perturbative QCD is not known. This is because the full path integration in QCD can not be done analytically (see section II). For this reason the full path integration in QCD is done numerically by using lattice QCD method. It is claimed in the lattice QCD studies that the mass of the hadron can be extracted by using the lattice QCD method. This, however, is not true which we will show in this paper.

Since the lattice QCD does the full path integration of quark and gluon fields in QCD, see section II, all that lattice QCD can calculate is the vacuum expectation of the non-perturbative correlation function of the type
\bea
<\Omega|{\bar \psi}(x_1)...\psi(x_n)|\Omega>
\label{llt}
\eea
where $\psi(x)$ is the quark field and $|\Omega>$ is the full interacting vacuum in QCD. However, since the path integration in lattice QCD involves the quark and gluon fields (but not hadron fields), the lattice QCD can not calculate the hadronic observable.

Since lattice QCD can numerically calculate the non-perturbative vacuum expectation $<\Omega|{\bar \psi}(x_1)...\psi(x_n)|\Omega>$ in QCD in eq. (\ref{llt}) but can not calculate the hadronic observable, the lattice QCD method inserts complete set of hadron states $\sum_n |n><n|=1$ in between appropriate partonic operators in eq. (\ref{llt}) to extract the hadronic observable [see eqs. (\ref{dxc}) and (\ref{gxc}) for details].

The time evolution of an operator ${\hat {\cal O}}(t)$ in the lattice QCD is governed by the Heisenberg evolution
\bea
{\hat {\cal O}}(t)=e^{-itH_{QCD}} {\hat {\cal O}} e^{itH_{QCD}}
\label{llt2}
\eea
where $H_{QCD}$ is the full QCD hamiltonian which includes all the quarks plus antiquarks plus gluons inside the hadron. One of the crucial assumption made by the lattice QCD method is
\bea
H_{QCD}|n> =E_n|n>
\label{exc}
\eea
where $|n>$ is the (physical) momentum eigenstate of the hadron normalized to unity and $E_n$ is the energy of the hadron which is a physical quantity.

However, in this paper we show that the eq. (\ref{exc}) is not correct due to confinement in QCD involving non-perturbative QCD, {\it i. e.}, we find
\bea
H_{QCD}|n> \neq E_n|n>.
\label{kxc}
\eea
This is because the QCD hamiltonian $H_{QCD}$ is unphysical but the $E_n$ and $|n>$ of the hadron are physical (see sections \ref{unphys} and \ref{phys} for details). We find that eq. (\ref{kxc}) is consistent with [see eq. (\ref{ncv}) for the derivation]
\bea
E_{\rm QCD}(t)=<n|H_{QCD}|n>\neq E_n
\label{hhj}
\eea
due to non-zero energy flux $E_{\rm flux}(t)$ in QCD because of confinement involving non-perturbative QCD. In eq. (\ref{hhj}) the $E_{QCD}(t)$ is the gauge invariant color singlet energy of all the quarks plus antiquarks plus gluons inside the hadron.

Hence we find that the lattice QCD method to study hadron mass is not correct.

The paper is organized as follows. In section II we briefly review the non-perturbative correlation function in QCD in the path integral formulation. In section III we discuss the lattice QCD method to study the hadron matrix element. In section IV we mention that the QCD hamiltonian and the QCD operator are not physical. In section V we mention that the momentum eigenstate and the mass/energy of the hadron are physical. In section VI we show that the unphysical QCD hamiltonian operating on physical eigenstate of hadron can not give the physical energy eigenvalue of the hadron. In section VII we show that the lattice QCD method to study hadron mass is not correct. Section VIII contains conclusions.

\section{ Non-perturbative Correlation Function Using Path Integral Formulation of QCD }

In order to extract the hadronic observable in the lattice QCD method one chooses an operator ${\hat {\cal O}}(x)$ built out of quark and antiquark fields in QCD which has the same quantum number of the hadron so that upon hadronization it generates the hadron. In the path integral formulation of the QCD the non-perturbative correlation function of the type $<\Omega|{\hat {\cal O}}(x_1)...{\hat {\cal O}}(x_n)|\Omega>$ is given by
\bea
&&<\Omega|~{\hat {\cal O}}(x_1)...{\hat {\cal O}}(x_n)|\Omega>=\nonumber \\
&&\frac{\int [d\psi][d{\bar \psi}][dA] ~{\rm det}[\frac{\delta \partial^\lambda A_\lambda^b}{\delta \omega^c}]~{\hat {\cal O}}(x_1)...{\hat {\cal O}}(x_n)~e^{-i\int d^4x [F_{\lambda \nu}^b(x)F^{\lambda \nu b}(x) -\frac{1}{2\alpha} (\partial^\lambda A_\lambda^b(x))^2+{\bar \psi}_j(x)[\delta^{jk}(i{\not \partial} -m)+gT^b_{jk}{\not A}^b(x)]\psi_k(x)]}}{\int [d\psi][d{\bar \psi}][dA] ~{\rm det}[\frac{\delta \partial^\lambda A_\lambda^b}{\delta \omega^c}]~e^{-i\int d^4x [F_{\lambda \nu}^b(x)F^{\lambda \nu b}(x) -\frac{1}{2\alpha} (\partial^\lambda A_\lambda^b(x))^2+{\bar \psi}_j(x)[\delta^{jk}(i{\not \partial} -m)+gT^b_{jk}{\not A}^b(x)]\psi_k(x)]}}\nonumber \\
\label{llt6}
\eea
where $\psi_i(x)$ is the quark field with color index $i=1,2,3$, the $A_\mu^a(x)$ is the (quantum) gluon field, $\alpha$ is the gauge fixing parameter and
\bea
F_{\nu \delta}^b(x) =\partial_\nu A_\delta^b(x) - \partial_\delta A_\nu^b(x)+gf^{bad} A_\nu^a(x) A_\delta^d(x).
\label{llt7}
\eea
There is no ghost field in eq. (\ref{llt6}) because we have used the determinant ${\rm det}[\frac{\delta \partial^\lambda A_\lambda^b}{\delta \omega^c}]$.

Due to the presence of cubic and quartic powers of the gluon field $A_\mu^a(x)$ in $F_{\lambda \nu}^b(x)F^{\lambda \nu b}(x)$ in eq. (\ref{llt6}) it is not possible to evaluate this full path integration in QCD in eq. (\ref{llt6}) analytically. For this reason the full path integration in QCD in eq. (\ref{llt6}) is evaluated numerically by using the lattice QCD method. Note that in order to perform the numerical integration in lattice QCD it is necessary to go to Euclidean time instead of Minkowski time.

\section{ Lattice QCD Method To Study Hadronic Matrix Element}

In the lattice QCD method the vacuum-to-hadron matrix element and the mass of the hadron are extracted from the non-perturbative two-point correlation function \cite{pdg}. Similarly in the lattice QCD method the hadron-to-hadron matrix element is extracted from the non-perturbative three-point correlation function \cite{pdg}.

Note that the quark and antiquark fields in the operator ${\hat {\cal O}}(r,t)$ are not free quark and antiquark fields but these quark and antiquark fields are in the presence of the gluon field. Hence the non-perturbative two-point correlation function of the type
\bea
{\cal M}_2=\sum_r<\Omega|{\hat {\cal O}}(r,t){\hat {\cal O}}(0)|\Omega>
\label{bxc}
\eea
is in the full interacting QCD which is evaluated numerically by using the lattice QCD method. In order to extract the vacuum-to-hadron matrix element in the lattice QCD method one inserts complete set of hadron states
\bea
\sum_n |n><n|=1
\label{cxc}
\eea
along with eq. (\ref{llt2}) in eq. (\ref{bxc}) to find in the Euclidean time
\bea
&& \sum_r<\Omega|{\hat {\cal O}}(r,t){\hat {\cal O}}(0)|\Omega>=\sum_n <\Omega|e^{tH_{QCD}}{\hat {\cal O}}e^{-tH_{QCD}}|n><n|{\hat {\cal O}}|\Omega>\nonumber \\
&&=\sum_n <\Omega|{\hat {\cal O}}|n><n|{\hat {\cal O}}|\Omega> e^{-tE_n}
\label{dxc}
\eea
where $H_{QCD}$ is the (unphysical) QCD hamiltonian of all the quarks plus antiquarks plus gluons inside the hadron, $|n>$ and $E_n$ are the momentum eigenstate and energy of the hadron respectively which are physical. The crucial assumption made in the lattice QCD method to obtain the right hand side of eq. (\ref{dxc}) is the use of eq. (\ref{exc}). From eq. (\ref{dxc}) one can extract the vacuum-to-hadron matrix element $<n|{\hat {\cal O}}|\Omega>$ in the lattice QCD method in the large Euclidean time limit.

Similarly in the lattice QCD method the hadron-to-hadron matrix element $<n|{\hat {\cal O}}|n'>$ can be extracted from the non-perturbative three-point correlation function of the type
\bea
{\cal M}_3=\sum_r\sum_{r'}<\Omega|{\hat {\cal O}}(r,t){\hat {\cal O}}(r',t'){\hat {\cal O}}(0)|\Omega>.
\label{fxc}
\eea
Using eqs. (\ref{cxc}) and (\ref{llt2}) in (\ref{fxc}) one finds in the Euclidean time
\bea
&& \sum_r\sum_{r'}<\Omega|{\hat {\cal O}}(r,t){\hat {\cal O}}(r',t'){\hat {\cal O}}(0)|\Omega>=\nonumber \\
&&\sum_n \sum_{n'} <\Omega|e^{tH_{QCD}}{\hat {\cal O}}e^{-tH_{QCD}}|n><n|e^{t'H_{QCD}}{\hat {\cal O}}e^{-t'H_{QCD}}|n'><n'|{\hat {\cal O}}|\Omega>\nonumber \\
&&=\sum_n \sum_{n'} <\Omega|{\hat {\cal O}}|n><n|{\hat {\cal O}}|n'><n'|{\hat {\cal O}}|\Omega> e^{-(t-t')E_n-t'E_{n'}}.
\label{gxc}
\eea
Using the extracted vacuum-to-hadron matrix element $<n|{\hat {\cal O}}|\Omega>$ from eq. (\ref{dxc}) in eq. (\ref{gxc}) one extracts the hadron-to-hadron matrix element $<n|{\hat {\cal O}}|n'>$ from eq. (\ref{gxc}) in the lattice QCD method in the large Euclidean time limit.

\section{ QCD Hamiltonian and QCD Operator Are Not Physical }\label{unphys}

Now let us turn our discussion to the serious physical problems in the eqs. (\ref{dxc}) and (\ref{gxc}) in the lattice QCD method.

First of all since the lattice QCD evaluates the path integration of quark, antiquark and gluon fields numerically it has no information about the hadron while evaluating the path integration in QCD. The lattice QCD uses the (unphysical) QCD hamiltonian and the (unphysical) QCD operators of the quark, antiquark and gluon fields, see eqs. (\ref{dxc}) and (\ref{gxc}). Even if the operator ${\hat {\cal O}}(r,t)$ used in the path integration in QCD in eq. (\ref{llt6}) has the same quantum numbers of the hadron but this operator ${\hat {\cal O}}(r,t)$ is not a hadronic operator. The operator ${\hat {\cal O}}(r,t)$ is still the partonic operator even if it is color singlet, gauge invariant and carries the same quantum numbers of the hadron. Hence one finds that all that lattice QCD numerically evaluates is the path integration of the quark, antiquark and gluon fields to predict the vacuum-to-vacuum expectation value of the non-perturbative correlation function of the type $<\Omega|{\hat {\cal O}}(x_1)...{\hat {\cal O}}(x_n)|\Omega>$.

In order to convert the vacuum-to-vacuum expectation value of the non-perturbative correlation function of the type $<\Omega|{\hat {\cal O}}(x_1)...{\hat {\cal O}}(x_n)|\Omega>$ to (physical) hadronic observables, the lattice QCD makes assumptions as described in eqs. (\ref{dxc}) and (\ref{gxc}). However, we find in this paper that there are serious physical problems in eqs. (\ref{dxc}) and (\ref{gxc}) in the lattice QCD method which we will discuss in this paper.

The major flaw in the lattice QCD method of inserting complete set of hadronic states in between partonic operators is in the use of eq. (\ref{exc}) in eqs. (\ref{dxc}) and (\ref{gxc}). According to eq. (\ref{exc}) the $H_{QCD}$ is the (unphysical) QCD hamiltonian of all the quarks plus antiquarks plus gluons inside the hadron. This QCD hamiltonian $H_{QCD}$ is not physical because we have not directly experimentally observed quarks and gluons. Similarly the operator ${\hat {\cal O}}(r,t)$ of the partons in eq. (\ref{llt6}) is not physical even if it is color singlet, gauge invariant and carries the same quantum numbers of the hadron because we have not directly experimentally observed quarks and gluons.

Because of this reason there is a serious physical problem in operating the unphysical QCD hamiltonian $H_{QCD}$ on the physical eigenstate $|n>$ of the hadron to obtain the physical energy eigenvalue $E_n$ of the hadron [see eq. (\ref{exc})] which we will discuss in detail in sections \ref{phys} and \ref{ncr}.

The energy $E_{QCD}(t)$ of all the quarks plus antiquarks plus gluons inside the hadron corresponding to this QCD hamiltonian $H_{QCD}$ is given by \cite{nkhm}
\bea
E_{QCD}(t)=<n|\int d^3x ~{\hat T}^{00}_{QCD}({\vec x},t)|n>= <n|H_{QCD}|n>\neq E_n
\label{ncv}
\eea
which is not a conserved quantity because of the non-vanishing energy flux $E_{\rm flux}(t)$ in QCD. The non-zero energy flux $E_{\rm flux}(t)$ in QCD arises due to the non-vanishing boundary surface term because the potential energy in QCD is an increasing function of distance due to confinement in QCD which involves non-perturbative QCD \cite{nkhm}. In eq. (\ref{ncv}) the ${\hat T}^{00}_{QCD}({\vec x},t)$ is the $00$ component of the energy-momentum tensor density operator ${\hat T}_{QCD}^{\mu \nu}(x)$ of all the quarks plus antiquarks plus gluons inside the hadron.

This is not surprising because the quark/antiquark and the gluon are not directly experimentally observed and hence the energy $E_{QCD}(t)$ of all the quarks plus antiquarks plus gluons inside the hadron is not a physical quantity.

\section{ Momentum Eigenstate and Mass/Energy of Hadron Are Physical }\label{phys}

Unlike the quark and gluon which are not directly experimentally observed, the hadron is directly experimentally observed. Because of this reason the quark, the antiquark and the gluon are not physical whereas the hadron is physical.

This implies that the QCD hamiltonian $H_{QCD}$ is unphysical because $H_{QCD}$ is the QCD hamiltonian of all the quarks plus antiquarks plus gluons inside the hadron. However, the momentum eigenstate $|n>$ of the hadron, the energy $E_n$ of the hadron and the mass $M_n$ of the hadron are physical.

\section{ Unphysical QCD Hamiltonian Operating On Physical Eigenstate of Hadron Can Not Give The Physical Energy Eigenvalue of Hadron }\label{ncr}

Note that everything in eq. (\ref{exc}) are physical except the QCD hamiltonian $H_{QCD}$. This implies that the eq. (\ref{exc}) can not be correct because the right hand side of eq. (\ref{exc}) is physical but the left hand side of eq. (\ref{exc}) is not physical due to the the presence of unphysical QCD hamiltonian $H_{QCD}$ in the left hand side of eq. (\ref{exc}).

We have shown in eq. (\ref{ncv}) that due to the confinement involving non-perturbative QCD the energy $E_{QCD}(t)$ of all the quarks plus antiquarks plus gluons inside the hadron is not a conserved quantity due to the non-vanishing energy flux $E_{\rm flux}(t)$ in QCD \cite{nkhm}. Hence unless $E_{QCD}(t)=E_n$ the eq. (\ref{exc}) is not satisfied.

However, since the energy $E_{QCD}(t)$ of all the quarks plus antiquarks plus gluons inside the hadron depends on time $t$ and the energy $E_n$ of the hadron is constant (independent of time $t$) we find that \cite{nkhm}
\bea
E_{QCD}(t)\neq E_n.
\label{jxc}
\eea
Using eq. (\ref{jxc}) in (\ref{hhj}) we derive eq. (\ref{kxc}).

Hence we find that in QCD the eq. (\ref{exc}) is not correct but the eq. (\ref{kxc}) is correct due to the existence of non-zero energy flux $E_{\rm flux}(t)$ in QCD \cite{nkhm} which is because of confinement involving non-perturbative QCD.

\section{ Lattice QCD Method to study hadron mass is not correct }

Consider the pion mass $M_\pi$ extraction using the lattice QCD method. For the hadron to be a pion the QCD operator ${\hat {\cal O}}(r,t)$ is given by
\bea
{\hat {\cal O}}(x) =\psi^\dagger(x) \gamma_5 \psi(x).
\label{axc}
\eea
In the non-perturbative two-point correlation function in QCD in eq. (\ref{bxc}) the operator ${\hat {\cal O}}(0)$ creates the quark-antiquark pair at the space-time point $0$ from the QCD vacuum with same quantum number of the pion. The evolution of this quark-antiquark state from the space-time point $0$ to the space-time point $({\vec r},t)$ is via the QCD hamiltonian $H_{QCD}$. This QCD hamiltonian $H_{QCD}$ is the hamiltonian of all the quarks plus antiquarks plus gluons inside the pion. Finally, the operator ${\hat {\cal O}}(r,t)$ annihilates this quark-antiquark pair.

Using the operator ${\hat {\cal O}}(x)$ from eq. (\ref{axc}) in eq. (\ref{dxc}) we find for the lowest energy state in the large Euclidean time limit \cite{pdg}
\bea
&& \sum_r<\Omega|{\hat {\cal O}}(r,t){\hat {\cal O}}(0)|\Omega>|_{t\rightarrow \infty} = <\Omega|{\hat {\cal O}}|\pi><\pi|{\hat {\cal O}}|\Omega> e^{-tM_\pi}
\label{hxc}
\eea
where
\bea
<\Omega|{\hat {\cal O}}|\pi>=f_\pi M_\pi
\label{ixc}
\eea
with $f_\pi$ being the pion decay constant. This is the usual procedure to extract the pion mass $M_\pi$ in the lattice QCD method.

Note that the eq. (\ref{exc}) is used in eq. (\ref{hxc}) to extract the pion mass $M_\pi$ in lattice QCD method. However, as we have shown in the section \ref{ncr} the eq. (\ref{exc}) is not correct in QCD but the eq. (\ref{kxc}) is correct in QCD due to the existence of non-zero energy flux $E_{\rm flux}(t)$ in QCD \cite{nkhm} which is because of confinement involving non-perturbative QCD.

Hence by using eq. (\ref{kxc}) in (\ref{dxc}) we find
\bea
&& \sum_r<\Omega|{\hat {\cal O}}(r,t){\hat {\cal O}}(0)|\Omega>=\sum_n <\Omega|e^{tH_{QCD}}{\hat {\cal O}}e^{-tH_{QCD}}|n><n|{\hat {\cal O}}|\Omega>\nonumber \\
&&\neq \sum_n <\Omega|{\hat {\cal O}}|n><n|{\hat {\cal O}}|\Omega> e^{-tE_n}
\label{lxc}
\eea
which gives [similar to eq. (\ref{hxc})]
\bea
&& \sum_r<\Omega|{\hat {\cal O}}(r,t){\hat {\cal O}}(0)|\Omega>|_{t\rightarrow \infty} \neq <\Omega|{\hat {\cal O}}|\pi><\pi|{\hat {\cal O}}|\Omega> e^{-tM_\pi}.
\label{mxc}
\eea
Hence we find from eq. (\ref{mxc}) that the lattice QCD method to extract the hadron mass is not correct.

Note that the hadronization also plays an important role to study the hadron production from the quark-gluon plasma at RHIC and LHC \cite{ln1,ln2,ln3,ln4}.

\section{Conclusions}
Since the numerical path integration in the lattice QCD involves quark and gluon fields (not hadron fields) the lattice QCD cannot calculate any hadronic observable. Because of this reason the hadronic properties are extracted in the lattice QCD method by inserting complete set of hadron states $\sum_n |n><n|=1$ in between the partonic operators by assuming $H_{\rm QCD}|n>=E_n|n>$ where $E_n$ is the energy of the hadron. However, in this paper we have found $H_{\rm QCD}|n>\neq E_n|n>$ because the QCD hamiltonian $H_{QCD}$ is unphysical but the $E_n$ and $|n>$ of the hadron are physical. We have shown that this is consistent with $E_{\rm QCD}(t)=<n|H_{QCD}|n>\neq E_n$ due to non-zero energy flux $E_{\rm flux}(t)$ in QCD because of confinement involving non-perturbative QCD. Hence we have found that the lattice QCD method to study hadron mass is not correct.

\end{document}